\documentclass[reprint]{revtex4-1}

\usepackage{bbold} 
\usepackage{graphicx}
\usepackage{verbatim}
\usepackage{mathtools}
\usepackage{url}
\usepackage{amsthm} 

\usepackage[final]{changes} 

\usepackage{hyperref} 

\usepackage{xfrac}
\usepackage{nicefrac}

\begin{document}

\title{Maxwell's demons with finite size and response time}

\author{Nathaniel Rupprecht, Dervis Can Vural}
\email{Corresponding Author: dvural@nd.edu}
\affiliation{University of Notre Dame}

\date{\today}

\begin{abstract}
Nearly all theoretical analyses of the Maxwell's demon focus on its energetic and entropic costs of operation. Here, we focus on its rate of operation. In our model, a demon's rate limitation stems from its finite response time and gate area. We determine the rate limits of mass and energy transfer, as well as entropic reduction for four such demons: Those that select particles according to (1) direction, (2) energy, (3) number and (4) entropy.
Lastly, we determine the optimal gate size for a demon with small, finite response time, and compare our predictions with molecular dynamics simulations with both ideal and non-ideal gasses. Lastly, we study the conditions under which the demons are able to move both energy and particles in the chosen direction when attempting to only move one.
\end{abstract}

\maketitle

A Maxwell's demon is a device that can measure the microstate of a closed system, thereby reducing its entropy, seemingly in violation with the second law \cite{maxwell1891theory}. A century of physics literature modelling the measurement protocols and internal workings of the demon \cite{thomson1874kinetic,smoluchowski1927experimentell,szilard1929entropieverminderung,brillouin1951maxwell} 
culminated to the conclusion that logically irreversible operations taking place within the demon \cite{bennett1982thermodynamics} such as information erasure \cite{landauer}, account for the lost entropy. 

Contemporary incarnations of the demon are capable of feedback control \cite{cao2004feedback,seifert2012stochastic} and universal computation \cite{zurek1999algorithmic,caves1990quantitative,caves1990comment}. Some demons measure and modify a tape of bits \cite{mandal2012work,mandal2013maxwell,barato2013autonomous,hosoya2015operational} or qubits \cite{deffner2013information} representing the state of a system, while others omit measurement altogether, instead, sorting through the microstates mechanically \cite{feynman2011feynman,gordon1983maxwell,skordos1992maxwell,tu2008efficiency}. Non-ideal demons have also been explored; \cite{mandal2012work,mandal2013maxwell} accounts for the thermal equilibriation of the demon with the system, and \cite{tu2008efficiency} studies the efficiency of an imperfect ratchet with finite mass. Today, we can build demons in the laboratory \cite{strasberg2013thermodynamics,roldan2014universal,cottet2017observing,thorn2008experimental,bannerman2009single,koski2015chip,vidrighin2016photonic,camati2016experimental,cottet2017observing,chida2017power,schaller2011probing,esposito2012stochastic,schaller2018electronic}, and even make practical use of them for harvesting energy \cite{gammaitoni2012there,bo2015thermodynamic,chida2017power,sothmann2014thermoelectric,roche2015harvesting,hartmann2015voltage}, or sorting atoms \cite{raizen2011demons}.

Experimentalists often discuss the temporal limitations of their demons, but nevertheless still operate under the simplifying assumption that the time $\tau$ it takes to sense, process and respond to information is negligible compared to all other times \cite{chida2017power,schaller2011probing,esposito2012stochastic,vidrighin2016photonic,schaller2018electronic,strasberg2013thermodynamics}. \added{While feedback control demons that operate periodically have been studied \cite{schaller2018electronic,engelhardt2018maxwell,bergli2013information}, much of their focus is on the \(\tau \to 0\) or \(\tau \to \infty\) limits, not on how the demon changes as \(\tau\) changes, and neither treat the limitations of the demons as their main object of study.}

\added{The optimality of resetting or erasing a single bit in finite time is well understood \cite{berut2012experimental,browne2014guaranteed,zulkowski2014optimal}. In many-body context,} cells constitute information engines that perform measurements and computations to process energy in a highly stochastic environment \cite{mehta2012energetic,barato2013information,lang2014thermodynamics,barato2014efficiency,sartori2014thermodynamic}. Here too, the timescale at which the cell operates relative to the time scale of its environment impacts its efficiency of information processing \cite{barato2014efficiency}.

In this paper, we study how the transport rate attainable by a Maxwell demon operating between two chambers of gas is restricted by the \added{finite area} \(A\) of the gate that the demon controls, and the rate \added{\(1/\tau\)} at which the demon operates. In practice, \(A\) and \(\tau\) would be constrained by experimental practicalities such as inertia and friction. Ultimately however, theoretical bounds on speed, length and mass set the true limits on how quickly a demon can transport energy or particles. For example, the gate cannot close faster than the speed of light, and must necessarily be larger than the thermal wavelength of an atom.  
 
To this end, we study four spatio-temporally limited demons that make decisions based on direction, number, energy or entropy measurements. For all four, we obtain heat, mass and entropy transport as a function of \(\tau,\, A\). 
We compare our results to molecular dynamics simulations, and study the conditions under which a demon is able to move both energy and particles from left to right when only aiming to move one or the other.
 
\section*{Problem setup}

Consider a partition separating two volumes of ideal gas, labeled as left (l) and right (r), with volumes \(V_l, V_r\), energies \(E_l,\,E_r\) and numbers \added{of particles} \(N_l,\,N_r\)

In the partition between the volumes is a gate of area \(A\), which the demon has control over. Except for the possibility of particles passing through the gate when it is open, the partitions are isolated from one another.

We also assume that each partition is large enough that it is a self averaging canonical distribution. In this case, the speed distribution for a particle is
\begin{align}
p(v) = \Omega_d v^{d-1} e^{-\beta_s m v^2/2}/Z_s,
\quad
Z_s = [ 2\pi/(m\beta_s)]^{d/2}\nonumber
\end{align}
where \(\Omega_d = 2,\,2\pi,\,4\pi\) is the solid angle in \(d=1,\,2,\,3\) dimensions and \(m\) is the particle mass. The index \(s \in \{l, r\}\) represents a generic side. The temperature, \(1/\beta_s\), and energy per particle \(\bar{E}_s =E_s/N_s\) are related by \(\beta_s E_s = N_s  d/2 \equiv \beta_s N_s \bar{E}_s\), by the equipartition theorem.

We assume that the demon decides on the state of the gate every \(\tau\) \added{seconds, which models all delays, e.g. due to measurement, processing or physical response. We assume that after every $\tau$, the state of the gate is updated instantaneously.} Since we are interested in determining how the physical limitations of the demon restrict its ability to operate separately from information theoretical restrictions, we are not concerned with how the demon acquires information, nor how it computes its decisions.

Let \(\hat{\mathcal{A}}\) be the event that a particle arrives at the gate within a duration of \(\tau\), (i.e. passing through it if it is open, or bouncing off it if it is closed). For a randomly chosen particle with given speed \(v\) the probability of \(\hat{\mathcal{A}}\) is \(p(\hat{\mathcal{A}} \vert v) = c_d v \tau A/V\), where \(c_d=1/2, 1/\pi, 1/4\) in \(d=1,2,3\) dimensions, (we define \(A\equiv 1\) for $d=1$). Thus, the probability that a random particle on side \(s\) impinges upon the gate during \added{the time window} is
\begin{align}
p(\hat{\mathcal{A}}) &= \int_0^\infty p(\hat{\mathcal{A}} \vert v)p(v) \mathbf{d}v = \kappa_s/N_s
\label{ProbabilityOfEscape}
\\
\kappa_s &= \frac{\rho_s \tau A}{\sqrt{2 \pi \beta_s m}} = \rho_s\tau A \sqrt{\frac{\bar{E}_s}{d\,\pi\,m}} \equiv \nu_s \tau .\label{kappa}
\end{align}
We will be interested in the thermodynamic limit, \(N_s,\,V_s,\,E_s \to \infty\) keeping \(\bar{E}_s \equiv E_s/N_s\) and \(\rho_s \equiv N_s/V_s\) constant. For estimates on the error that this introduces for finite systems, see \ref{Appendix:CalcDetails}.

Knowing the probability that a random particle with a specific velocity arrives at the gate allows us to compute the probability that exactly \(n\) particles carrying total energy \(E\) arrives at the gate during \added{within a duration} \(\tau\) (see \ref{Appendix:CalcDetails}),
\begin{align}
p(E,n) &= \frac{\kappa_s^n (\beta_s E)^{nD} }{\Gamma(nD)n!} \frac{e^{-\beta_s E-\kappa_s}}{E}
\label{ProbabilityNE}
\end{align}
which can be marginalized over number or energy to find the probability of number and the probability of energy,
\begin{align}
p(\added{n \vert } n>0) &= \frac{\kappa^n}{n!}e^{-\kappa} , \quad p(n=0) = e^{-\kappa} \delta(E)
\label{ProbabilityOfNumber}
\\
p(E) &= \frac{1}{E} e^{-\beta E-\kappa} \sum_{n=1}^\infty \frac{\kappa^n(\beta E)^{nD}}{n! \Gamma(nD)}
\label{ProbabilityOfEnergy}
\end{align}
where \(D=(d+1)/2\).  
The incomplete energy moments can be found in terms of incomplete gamma functions, \(\Gamma(\cdot, \cdot)\),
\begin{align}
\langle E^s \rangle_{\ge E_0}\! = \!\!\int_{E_0}^\infty \!\!\!\!E^s p(E) = \frac{e^{-\kappa}}{\beta^s} \sum_{n=1}^\infty \frac{\kappa^n}{n!} \frac{\Gamma \left(nD+s, \beta E_0\right)}{\Gamma(nD)}\nonumber
\end{align}
For complete energy moments, the incomplete gamma function is replaced with a gamma function. For \(s=1\) we get the average \(\langle E \rangle = \kappa D/\beta\).
The number distribution moments can be found similarly, \(\langle n^s \rangle = e^{-\kappa} \left( \kappa \, \partial_\kappa \right)^s e^\kappa\).

\subsection*{Entropy reduction by a demon}

Differentiating the Sackur-Tetrode equation with respect to time, we can find the entropy rates of the subsytems in terms of \(\dot{N}\), \(\dot{E}\). Adding the entropy rates for the two subsystems, and using
mass and energy conservation, \(\!\dot{N}_l = \!\!-\dot{N}_r\), \(\dot{E}_l = \!\!-\dot{E}_r\), we find that the change in entropy of the whole system is
\begin{align}
\frac{\dot{S}_\mathrm{tot}}{k_B} &= (\beta_r - \beta_l)  P_\tau + \left( \frac{d}{2} \log \left( \frac{\beta_l}{\beta_r}\right) - \log \frac{\rho_r}{\rho_l} \right)  I_\tau
\end{align}
where \( I_\tau\) and \( P_\tau\) are the number and energy currents. Note that the same answer is obtained when differentiating the purely classical \added{Clausius} entropy.
For more details on demon \added{entropy}, and notes on the entropic cost of operating the demons, see \ref{Appendix:DemonEntropy}.

\begin{figure*}
    \centering
    \includegraphics[width=0.97\textwidth]{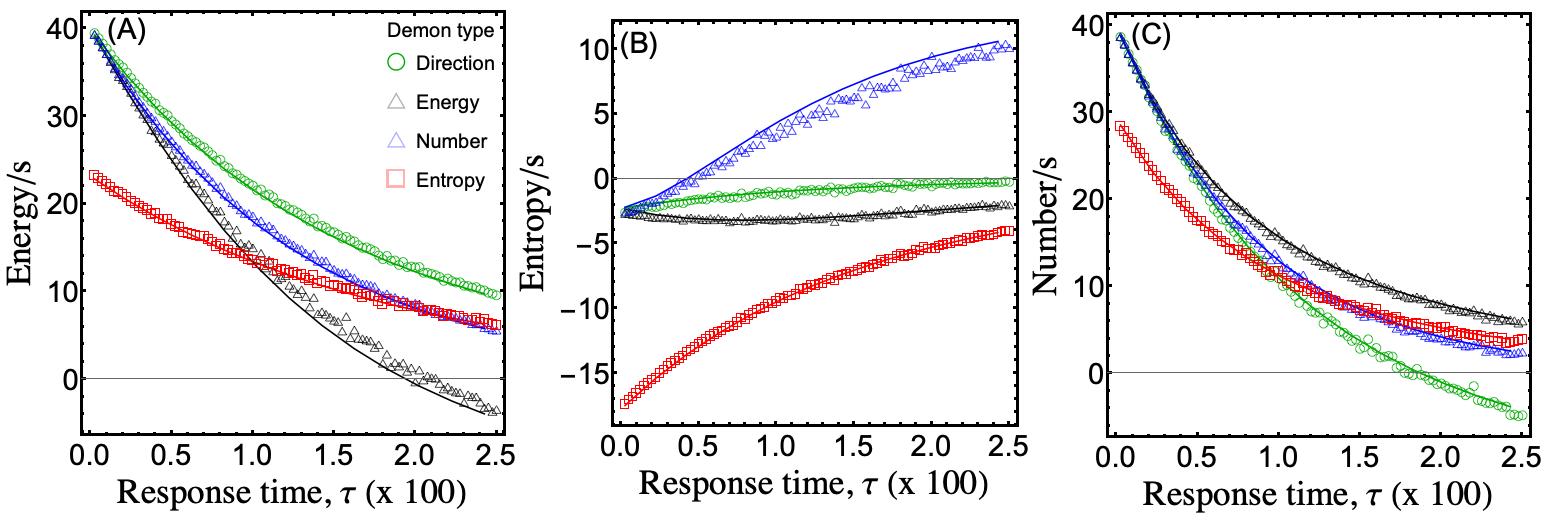}
    \caption{\textbf{Demon heat and mass transfer rates.} Two different one dimensional systems, with heat and mass currents plotted as functions of the demon's temporal resolution \(\tau\). Theory (solid lines) agrees with simulations (dots). \textbf{(A)}: The subsystems have a temperature difference, but with the same number density, \(\beta_l = 1\), \(\beta_r = 0.25\), \(\rho_l = \rho_r = 100\). Note that the heat transfer rate can be negative for the number demon. \textbf{(B):} The change in system entropy per unit time for the same system as in the left panel. \textbf{(C):} The right subsystem has a \emph{lower} temperature, and a much higher number density than the left subsystem, \(\rho_l = 100,  \rho_r = 400\), \(\beta_l=1, \beta_r=2\). Here, it can be seen that the energy demon has a negative number current for large \(\tau\).}
    \label{Fig:RatesVsTimeFigure}
\end{figure*}

\section*{Demon Models}
 We model four demons who make decisions based on direction, energy, number and entropy. Our convention will be that each demon will attempt to move its target quantity, e.g. mass, heat, from the left partition to the right partition. For all four demons, we calculate heat \( P_\tau\), number \( I_\tau\), and entropy \( J_\tau\) currents as a function of the gate area and response time, and compare these with Monte Carlo simulations (Fig. \ref{Fig:RatesVsTimeFigure}). We assume that even though the demon is operating, the two volumes each remain in equilibrium.

{\textbf{(1)} A \textbf{direction demon}} opens the gate only if there are no particles moving from right to left. Since the probability that no particles approach from the right is \(e^{-\kappa_r}\), the average energy approaching the gate from the left is \(D  \kappa_l/\beta_l\), and the average number approaching the gate from the left is \(\kappa_l\), the average heat and mass currents are
\begin{align}
 P_\tau^{(d)} = D \nu_l e^{-\kappa_r}/\beta_l,
\qquad
 I_\tau^{(d)} = \nu_l e^{-\kappa_r}
\end{align}

Thus, the performance of the demon falls exponentially with \(\tau\). For an infinitely precise demon that can process all incoming particles (\(\tau \to 0\)), the rate of heat transfer is 
\( P^{(s)}_0 = (2 \bar{E}_l\rho_l D A) \sqrt{\bar{E}_l/(d^3 \pi m)}\).
Naturally, this only depends on the left subsystem, since the demon can shut out the right subsystem completely.

Interestingly, there is an optimal value for the area of the gate. Optimizing \(A\) for fixed \(\tau\) and other system parameters shows that \(A^* = (\tau\rho_r)^{-1} \sqrt{d \pi m/\bar{E}_r}\) maximizes both mass and heat transport, for \(d>1\). 

{\textbf{(2)} An \textbf{energy demon}} opens the gate whenever the right moving particles have greater energy than left moving ones. The energy demon's heat and mass transport rate converges to that of the direction demon as \(\kappa \to 0\), since the probability that multiple particles approach the gate from the right and left simultaneously, vanishes. Therefore, we can write the energy demon's heat and mass currents as the direction demon's, plus correction terms (for an example derivation, see \ref{Appendix:HigherDimensionDemons}).
\begin{align}
     P_\tau^{(e)} \!&=  P_\tau^{(d)} + \frac{\nu_l \nu_r\Gamma(2D+1)}{\Gamma(D)^2} (-1)^D \left( \frac{f_1}{\beta_l} + \frac{f_2}{\beta_r} \right) \tau\, e^{-\kappa_l-\kappa_r}
     \nonumber \\
     I_\tau^{(e)} \!&=  I^{(d)}_\tau + \frac{\Gamma(3D)\,\tau^2\,e^{-\nu_l - \kappa_r}}{2\,\Gamma(2D) \Gamma(D+1)} \left( \nu_l^2 \nu_r f_3 - \frac{\nu_l \nu_r^2}{2} f_4 \right)
    \label{PowerDemonCurrents}
\end{align}
where \(f_1 = B(-\beta_l/\beta_r, D, -2D)\) and \(f_2 = B(-\beta_l/\beta_r, D+1, -2D)\) are Euler beta functions, and
\(f_3=F_{2,1} (D, 3D; D+1; - \beta_r/\beta_l)\), and \(f_4 = F_{2,1}(2D, 3D; 2D+1; -\beta_l/\beta_r)\) are hypergeometric functions. 

It is not difficult to numerically solve for \(P_\tau^{(e)},\,I_\tau^{(e)}\) for higher dimensions. An exact analytical solution for heat and mass transport for \(d=1\) is given in  \ref{Appendix:DemonSolutions}.

\textbf{(3)} A \textbf{number demon} opens the gate if right-moving particles are more than left-moving ones. An exact solution for \(d=1\) is given in \ref{Appendix:DemonSolutions}, for \(d>1\), we again obtain the leading order correction,
\begin{align}
 P_\tau^{(n)} &=  P^{(d)}_\tau + \frac{\nu_l^2 \nu_r}{2 \beta_l} \tau^2\,(d+2) \left[ 1 - \frac{1}{2} \frac{d+3}{d+2} \frac{\beta_l}{\beta_r} \right] e^{-\kappa_l-\kappa_r}\nonumber
\\
 I_\tau^{(n)} &=  I^{(d)}_\tau +  \frac{\nu_l^2 \nu_r}{2}\tau^2\, e^{-\kappa_l-\kappa_r}
\label{NumberDemonCurrents}
\end{align}

\textbf{(4)} An \textbf{entropy demon} opens the gate if doing so reduces the total entropy, i.e. if
\begin{align}
    E_r - E_l > \left[ \log \frac{\rho_r}{\rho_l} - \frac{d}{2} \log \left( \frac{\beta_l}{\beta_r}\right)\right]\frac{n_l - n_r}{\beta_l-\beta_r}\equiv\chi \cdot(n_l - n_r)\nonumber
\end{align}

If \(\beta_l=\beta_r\), the entropy demon opens the gate whenever \(n_l > n_r\), acting as a number demon, and if \(\chi=0\), it acts as an energy demon. The average heat and mass flow is
\begin{align*}
    & J = \sum_{n_l, n_r=0}^{\infty} \int_0^\infty \mathbf{d}E_l\,\mathbf{d}E_r\, p_{n_l}^{(l)}(E_l) \,  p_{n_r}^{(r)}(E_r) j(\{E_s\}, \{n_s\})
    \\
    & j(\{E_s\}, \{n_s\}) = \Theta \left(E_r-E_l-\chi(n_l - n_r)\right) \Delta(\{E_s\}, \{n_s\})
\end{align*}
with a step function enforcing the inequality above. Here \(\Delta=E_l-E_r\) for \( J={P_\tau}\) and \(\Delta=n_l-n_r\) for \(J={I_\tau}\).

The entropy demon behaves different than the number and energy demons, it does not act as a direction demon as \(\tau \to 0\).
\begin{figure}
    \centering
    \includegraphics[width=0.49\textwidth]{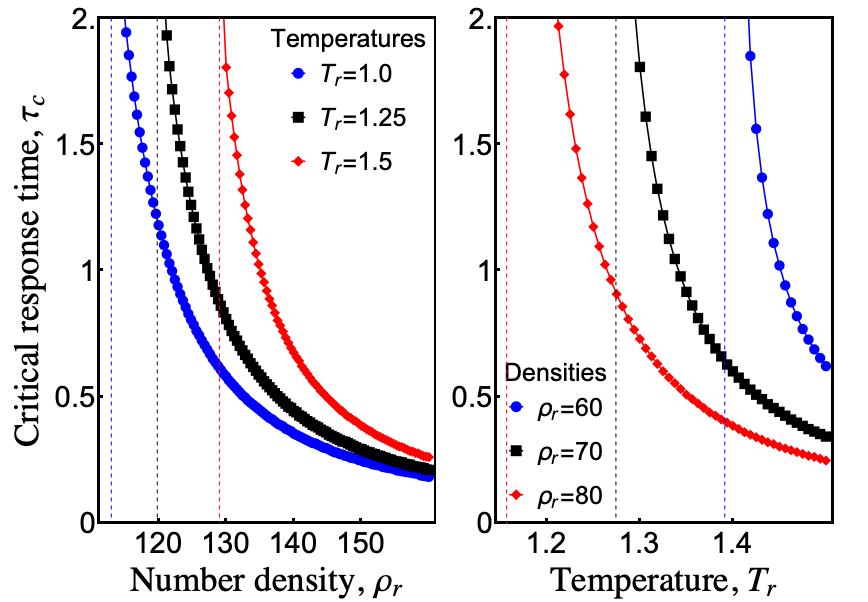}
    \caption{\textbf{Critical response time.} The critical \(\tau\) for energy and number demons, along with estimates of the critical values below which no \(\tau_c\) exists. Above \(\tau_c\), the demon will transfer heat or mass from right to left instead of from left to right. In both plots, the left subsystem has \(T_l=1\), \(\rho_l=100\), only the parameters of the right subsystem are varied. \textbf{Left:} The energy demon's \(\tau_c\) for different right subsystem temperatures, varying right subsystem number density. \textbf{Right:} The number demon's \(\tau_c\), for different right subsystem number densities, varying the right subsystem's temperature.}
    \label{Fig:TauCriticalFigure}
\end{figure}

\section*{Simulations}
We distribute particles uniformly in space, assign them Boltzmann-distributed velocities, and obtain the time they approach the gate and the energy they carry. 
For \(d>1\), the probability of atoms arriving the gate decreases with decreasing gate area. Thus, for economical reasons, we run most simulations only for \(d=1\). In all plots the units of temperature is such that \(k_B = 1\). See \ref{Appendix:SimulationDetails} for simulation details.

In Fig. \ref{Fig:RatesVsTimeFigure} we compare the energy, mass and entropy currents generated by the demons to our formulas. In panels A, B, the right chamber has a temperature four times greater than the left, and the number densities are the same. In panel C, the temperatures of the right chamber is half of that of the left, but the number density of the right chamber is four times that of the left.

\begin{figure}
    \centering
    \includegraphics[width=0.5\textwidth]{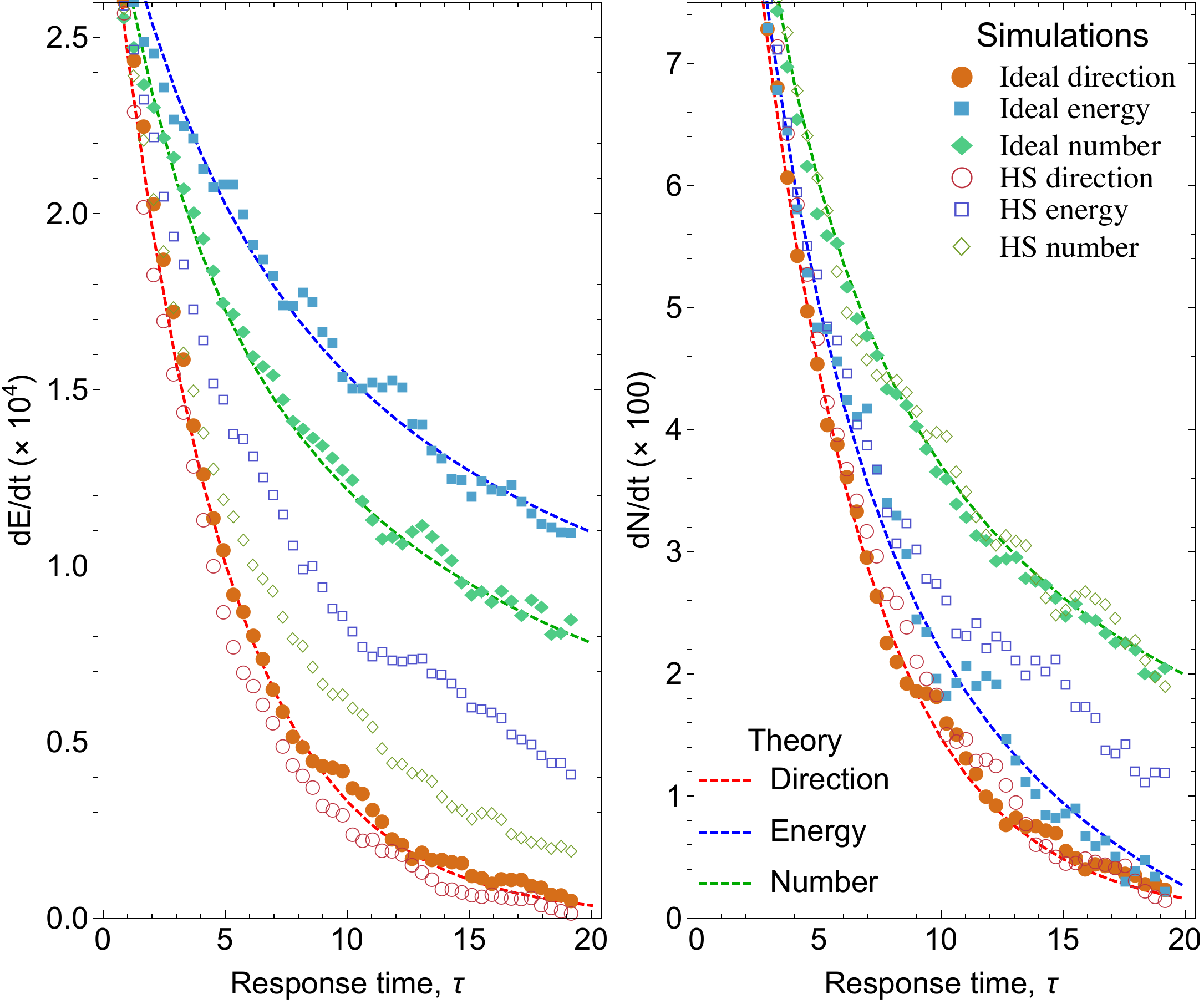}
    \caption{\textbf{Two dimensional demon.} A two dimensional maxwell demon, operating on an ideal gas and a hard sphere gas.  The energy (left) and number (right) rate of demons working with an ideal gas match well with prediction for all three demon types, and the hard sphere gas is qualitatively similar. The system has \(\rho_l = 1.5,\,\rho_r = 3\) and \(T_l = 0.0015,\,T_r = 0.001\). The dimensions of each subsystem are \(30 \times 30\), and the particle radius is 0.025. The packing fraction of the left and right subsystems are \(\phi_l = 0.0029\), \(\phi_r = 0.0059\).
    }
    \label{Fig:Demon2D}
\end{figure}

Fig. \ref{Fig:RatesVsTimeFigure} illustrates an interesting phenomenon. \added{For demons with fast response time (i.e. small \(\tau\)),} regardless of whether they are aiming to transport heat or mass, end up transporting both quantities in the same direction. However, at sufficiently large \(\tau\) (and, as we will see, large enough values of \(\rho_r\) or \(T_r\)), \added{heat and mass transport can be in opposite directions}. For example, the number demon is willing to let a few very energetic molecules move from right to left as long as a larger number of less energetic molecules move from left to right. 
We define \(\tau_c\) to be the response time for which demons start pumping particles or energy from right to left instead of left to right.

Fig. \ref{Fig:TauCriticalFigure} illustrates the behavior of \(\tau_c\). Theoretical values of \(\tau_c\) are plotted, varying either number density (for the energy demon, left), or temperature (for the number demon, right). 

For low enough number density or temperature, there may not be a \(\tau_c\) (it diverges at some critical number density or temperature). 
For \(\tau<\tau_c\), either demon strategy is appropriate for ensuring that there is no ``backwash'' of particles or energy from right to left, while above \(\tau_c\), a specific strategy must be favored to ensure this.
This illustrates another difference between our demons and ideal demons. An ideal, infinitely quick demon does not have to prioritize particle number or energy no matter what the number densities or temperatures of the subsystems are, but a restricted demon has to consider tradeoffs.

To check the generality of our prediction, we also ran molecular dynamics simulations of demons operating in two dimensions with ideal and hard sphere gasses. Due to computational constraints, the data is more noisy than the 1D case, but clearly the prediction and theory match well for ideal gas demons. Although our equations are only valid for ideal gasses, the currents for demons working with hard sphere gasses have qualitatively similar  decaying behavior as the predictions. For example, \(P^{(e)}\) and \(P^{(n)}\) are overestimated, and \(N^{(e)}\) is underestimated by the predictions, but all curves are qualitatively similar. We have also observed that the hard sphere demon's rates approach the corresponding ideal demon's rates as we reduce the volume of each individual particle. A video showing the demon in operation can be seen in the supplementary information. See \ref{Appendix:SimulationDetails} for more details on the simulation.

\section*{Discussion}
Most literature on the Maxwell's demon focuses on its thermodynamic \emph{cost of operation}. Here we point out that even if a demon has no restrictions on memory, or knowledge of the state of the systems, it
will still be limited in its \emph{rate of operation} due to its physical characteristics. Here, we determined rate bounds for four kinds of demons. We have derived the optimal area of the gate for the simple demon, and by extension, for all demons with small \(\tau\), and how the demons' response time and gate size determine heat, mass and entropy currents.

For a square gate with \(A=1\,\mu m^2\) that moves at the speed of light to sort air molecules at 300 K and standard pressure, we get \(\kappa \sim 9.5\). For a simple demon, the energy and number transfer for a demon with \(\tau>0\) is \(e^{-\kappa_r}\) times less than a demon operating with \(\tau=0\), meaning that its currents would be \(\sim 7.5 \times 10^{-5}\) times less than an infinitely fast demon.

Of course, not all realizations of Maxwell's demons operate via gates. For example, many nano-molecular pumps and refrigerators are implemented by single electron transistors. However, these devices still have spatial and temporal restrictions that play similar roles, such as finite sampling and feedback rates, the probability rate of electron tunneling and co-tunneling events, and quantum confinement effects \cite{schaller2011probing,deng2007compact,deng2007compactP2}. As such devices become widespread, it will be crucial to know how their spatial and temporal limitations influence transport rates.

\onecolumngrid

\setcounter{section}{0}
\renewcommand{\thesubsection}{Appendix \Alph{subsection}}

\subsection{Calculation details} \label{Appendix:CalcDetails}

In this appendix, we give more detailed derivations of the probabilities, along with detailed assumptions and estimates of the error associated with them.

\textbf{Arrival probabilities and error bounds:} First, we derive the expression \(p(\hat{\mathcal{A}} \vert v) = c_d\, v\,\tau\,A/V\) for \(d=2,\,3\) (the one dimensional case is trivial). We work with a generic side of the partition, so we drop the \(l\), \(r\) subscripts from quantities such as \(\kappa,\,V,\) etc. A particle will only impinge upon the gate if its velocity vector is pointed at the gate, and if its distance from the point of intersection of the particle's trajectory and the gate is at a distance less than \(v\,\tau\) from the current position of the particle. Divide the area into small (infintessimal) regions of area \(da\). The differential volume of physical space for a point at a distance \(r\) from \(da\), and making an angle \(\theta\) with the vertical is \(r\,dr\,d\theta\) in two dimensions, and \(r^2\,\sin \theta\,dr\,d\theta\,d\phi\) in three dimensions, where \(\phi\) is the azimuthal angle. The volume of angle space that the particle at \(\{r,\,\theta\}\) in 2D, or \(\{r, \,\theta,\,\phi\}\) in 3D, can have where its velocity vector points to the volume \(da\) is the differential angle (or solid angle) \(d\varphi = \frac{\cos \theta \, da}{\Omega_d\,r^{d-1}}\). Combining these yields the probability density that a particle located at a position, relative to the differential area \(da\), will arrive at the area within \(\tau\) given that it has velocity \(v\),
\begin{align*}
    dp_2(\hat{\mathcal{A}} \, \vert\, r,\,\theta,\,v) &= \frac{\cos \theta}{2\,\pi} \Xi (v\,\tau - r)\, dr\,d\theta\,da
    \\
    dp_3(\hat{\mathcal{A}} \, \vert \,r,\,\theta,\,v) &= \frac{\sin \theta\,\cos \theta}{4\,\pi} \Xi (v\,\tau - r)\, dr\,d\theta\,da
\end{align*}
where the clamp function, \(\Xi\), is \(\Xi(x) = 0\) for \(x<0\) and \(\Xi(x) = x\) for \(x\ge 0\). The clamp function is necessary because only particles that are fast enough and close enough can impinge upon the gate area within \added{a duration} \(\tau\).

Integrating over all space, we obtain \(p_d(\hat{\mathcal{A}}\,\vert v)\),
\begin{align*}
    p_2(\hat{\mathcal{A}}\,\vert v) &= \frac{1}{2\,\pi\,V} \int_0^A da \, \int_{-\pi}^{\pi} d\theta\, \cos \theta \, \int_0^{v\,\tau} dr = \frac{1}{\pi} \frac{v\,\tau\,A}{V} \equiv c_2\,v\,\tau\,A/V
    \\
    p_3(\hat{\mathcal{A}}\,\vert v) &= \frac{1}{4\,\pi\,V} \int_0^A da \, \int_{-\pi}^{\pi} d\theta\, \sin \theta \, \cos \theta \, \int_0^{2\pi} d\phi \, \int_0^{v\,\tau} dr = \frac{1}{4} \frac{v\,\tau\,A}{V} \equiv c_3\,v\,\tau\,A/V.
\end{align*}
Note that we have made the assumption that the hemisphere of radius \(v \tau\) centered at any point in the gate lies completely within the volume of gas.

If the smallest hemisphere centered at a point in the gate that does not lie within the volume of gas has radius \(r_*\), then \(p(\hat{\mathcal{A}}\, \vert\, v)\) is certainly valid for velocities less than or equal to \(v_* \equiv r_*/\tau\). This introduces an error in the arrival probability (\ref{ProbabilityOfEscape}), which can be bounded,
\begin{align*}
   &p(\hat{\mathcal{A}}) = \int_0^\infty p(\hat{\mathcal{A}}\,\vert\,v) p(v)\, dv \le \int_0^{v_*} p(\hat{\mathcal{A}}\,\vert\,v) p(v)\, dv = \kappa/N - \int_{v_*}^\infty p(\hat{\mathcal{A}}\,\vert\,v) p(v)\, dv
    \\
    &\int_{v_*}^\infty p(\hat{\mathcal{A}}\,\vert\,v) p(v)\, dv = \frac{\kappa}{N} \frac{\Gamma \left(\frac{d+1}{2}, \frac{\beta}{2}m v_*^2 \right)}{\Gamma \left(\frac{d+1}{2} \right)}.
\end{align*}
Therefore, the error is 
\begin{align}
    \vert p_{\text{actual}}(\hat{\mathcal{A}}) - p_{\text{approx}}(\hat{\mathcal{A}}) \vert \le \frac{\kappa}{N} \frac{\Gamma \left(\frac{d+1}{2}, \frac{\beta}{2}m (r_*/\tau)^2 \right)}{\Gamma \left(\frac{d+1}{2} \right)} \sim  \frac{\kappa}{N\,\Gamma \left(\frac{d+1}{2}\right)} \left(\frac{\beta m v_*^2}{2}\right)^{\frac{d-1}{2}} \exp \left(-\frac{\beta}{2} m v_*^2 \right) \text{ as } v_* \to \infty
    \label{Appendix:PHitError}
\end{align}
and is clearly very close to zero for large \(r_*\). When we say that we assume that the partition of gas is large, what we mean is that we assume that \(v_* = r_* /\tau\) is large enough that (\ref{Appendix:PHitError}) is small.

One useful identity concerning the constants \(c_d\) which is used in deriving (\ref{ProbabilityOfEscape}) is 
\begin{align*}
c_d \Omega_d \Gamma \left(\frac{d+1}{2} \right) = \pi^{(d+1)/2}
\end{align*}
where \(\Omega_d=2,\,2\pi,\,4\pi\) and \(c_d=1/2, 1/\pi,1/4\) in \(d=1,\,2,\,3\) dimensions.

\textbf{Probability of energy and \added{particle} number:} The probability that exactly \(n\) \added{particles} carrying total energy \(E\) will arrive at the gate during \added{a period of time} \(\tau\) can be derived by first calculating the probability that \added{the} event \(\hat{\mathcal{A}}\) occurs for \added{exactly} \(n\) particles (call this \added{new} event \(\hat{\mathcal{A}}_n\)), \added{and the \(n\) particles have} velocities \(v_1,v_2,\dots,v_n\), 
\begin{align*}
p_n(\hat{\mathcal{A}_n}, v_1, \dots, v_n) &= {N \choose n} p(\hat{\mathcal{A}}, v_1) \dots p(\hat{\mathcal{A}}, v_n) p^c(\hat{\mathcal{A}})^{N-n} \nonumber
\\
&= {N \choose n} \left( \frac{c_d \tau A \Omega_d}{V Z_\beta} \right)^n v_1^d\cdot \dots \cdot v_n^d \cdot e^{-\kappa} \exp \left(-\frac{\beta}{2} m (v_1^2 + \dots + v_n^2)\right).
\end{align*}
Note that \(p(\hat{\mathcal{A}}, v) = p(\hat{\mathcal{A}} \vert v) \, p(v) \), and \(p^c(\hat{\mathcal{A}}) = 1-p(\hat{\mathcal{A}})\). From this, we can integrate over the \(v_k\) using a delta function to ensure that the kinetic energy is \(E\),
\begin{align*}
p(E,n) = \int_0^\infty \mathbf{d}v_1\dots \mathbf{d}v_n \, \delta\left( \frac{1}{2}m(v_1^2+\dots+v_n^2) - E \right) p_n(\hat{\mathcal{A}}_n, v_1, \dots, v_n)
\end{align*}
This can be done using an integral representation of the delta function, and using the formula
\begin{align*}
\frac{1}{2 \pi} \int_{-\infty}^\infty \frac{e^{i k E}}{(\beta+i k)^n} \mathbf{d}k = \frac{E^{n-1} e^{-\beta E}}{\Gamma(n)}.
\end{align*}

In the thermodynamic limit, \(p^c(\hat{\mathcal{A}})^N = (1-\frac{\kappa}{N})^N \to e^{-\kappa}\), \(p^c(\hat{\mathcal{A}})^{-n} \to 1\), and the product \({N \choose n} V^{-n} = \frac{N}{V} \frac{N-1}{V} \dots \frac{N-n+1}{V}\frac{1}{n!} \to \rho^n/n!\). The remainder of the constants to the \(n\)-th power, and the \(E^{n-1} = E^n/E\) become \(\kappa^n/E\). This leaves us with the expression for \(p(E,n)\), (\ref{ProbabilityNE}), as desired.

\textbf{Incorporating chemical potentials:} If there are chemical potentials, some changes must be made to the probabilities. Let us assume that \(\mu_r = \mu > 0\), \(\mu_l=0\) (only the difference in the \(\mu\)s will matter). Since all particles that arrive at the gate from the right will be able to pass through the door, the previously derived formula for \(p(E, n)\) is still valid. For the left side though, particles can only pass through the gate \added{(even if it is open)} if they have \(K \ge \mu \), meaning that they have \(v \ge v_{min} \equiv \sqrt{2 \mu/m}\). The probabilities of arriving, \(p(\hat{\mathcal{A}})\) and \(p(\hat{\mathcal{A}} \vert v)\), as derived before, are just the probability of arriving at the door area. To describe whether the particle arrives at the gate, and will be able to pass through if the gate is open, they must be amended
\begin{align*}
    p^\prime(\hat{\mathcal{A}}) &= \int_{v_{min}}^\infty p(\hat{\mathcal{A}} \vert v) p(v) \mathbf{d}v = \kappa(v_{min}) / N
    \\
    \kappa(v_{min}) &= \frac{\rho\,\tau\,A}{\sqrt{2\,\pi\,\beta\,m}} \times \left( \frac{m\,\beta}{2}\right)^{\frac{d+1}{2}} \frac{v_{min}^{d+1}\,E_{(d-1)/2}(\frac{\beta}{2} m\,v_{min})}{\Gamma(D)}.
\end{align*}
The \(E_n(z)\) is the exponential integral function. Since \(v_{min}^{d+1} E_{(d-1)/2}(\beta m v_{min}/2) \to \Gamma(D) (m \beta/2)^{(d+1)/2}\) as \(v_{min}\to 0\), this equation reproduces (\ref{kappa}) when \(\mu = 0\).

The expression for \(p_n\) for particles on the left side must be replaced with
\begin{align*}
    p^\prime_n(\hat{\mathcal{A}}_n,v_1,\dots,v_n) = {N \choose n} p(\hat{\mathcal{A}}, v_1) \dots p(\hat{\mathcal{A}}, v_n) \left( 1-p^\prime(\hat{\mathcal{A}}) \right)^{N-n} \Theta(v_1 - v_{min}) \dots \Theta(v_n - v_{min}).
\end{align*}
Integrating this over all velocities with the delta function constraining the energy will result in the new expression \(p^\prime(E, n)\), analogous to (\ref{ProbabilityNE}). This in turn can be marginalized and results in analoges to (\ref{ProbabilityOfEnergy}) and (\ref{ProbabilityOfNumber}).

A final change that must be made: whenever particles move from right to left, they gain kinetic energy \(\mu\). Because of this, the energy demon must not make sure that \(E_l > E_r\), but instead it must open the gate only when \(E_l > E_r + n_r \mu\).

\subsection{One dimensional demon solutions} \label{Appendix:DemonSolutions}

As seen in (\ref{PowerDemonCurrents}), etc., the expressions for the power and number rate of the demons are very complicated in general, even in the small \(\tau\) limit. For the one dimensional case, we have solved for the power and number transfer rates for the power and number demons. The solution is straightforward, though tedious, to derive.

For the energy demon in one dimension, 
\begin{align}
 P^{(p)}_\tau &=  \frac{\kappa_l e^{-\kappa_l}}{\beta_l \tau} \left( 1 + \gamma_r   \frac{\kappa_r}{\kappa_l}   e^{- \kappa_r} \sum_{j,k \ge 0} {k+j \choose k} \frac{(\gamma_l \kappa_l)^j (\gamma_r \kappa_r)^k}{(k+1)!} \left[ \mathcal{E}_{1,j}(\kappa_l) - \gamma_l \kappa_l  (k+j+1)  \mathcal{E}_{1,j+2}(\kappa_l) \right] \right)
\label{PowerDemonPowerOneD}
\\
 I^{(e)}_\tau &= \frac{\kappa_l}{\tau} e^{-\kappa_r} \left(1 + \kappa_r \gamma_r e^{-\kappa_l} \sum_{j,k\ge0}^\infty \sum_{l=0}^{j} {l+k+1 \choose k} (j-k) \frac{\kappa_l^j}{(j+1)!}  \frac{(\kappa_r \gamma_r)^k}{(k+1)!}   \gamma_l^l  \right).
\label{PowerDemonNumberOneD}
\end{align}
For the number demon in one dimension,
\begin{align}
 I^{(e)}_\tau &= \frac{\kappa_l e^{-\kappa_l-\kappa_r}}{\tau} \sum_{n\ge0} \frac{(\kappa_l \kappa_r)^n}{n!} \mathcal{E}^\prime_{1,n+1}(\kappa_l) = \frac{e^{-\kappa_l-\kappa_r}}{\tau} \sum_{ k \ge 1} k \left(\frac{\kappa_l}{\kappa_r}\right)^{k/2} I_k \left(2\sqrt{\kappa_l \kappa_r)}\right).
\label{NumberDemonNumberOneD}
\\
 P^{(n)}_\tau &= \frac{\kappa_l e^{-\kappa_r}}{\beta_l \tau} \left( 1 + \kappa_l \kappa_r e^{-\kappa_l}\sum_{j,k\ge 0} \frac{\kappa_l^j}{(j+k+2)!} \frac{(\kappa_l \kappa_r)^k}{(k+1)!} \left[ j+k+2 - (k+1)\frac{\beta_l}{\beta_r}\right]\right).
\label{NumberDemonPowerOneD}
\end{align}
Here, \(\mathcal{E}_{\alpha,\beta}(z) = \sum_{k=0}^\infty \frac{z^k}{\Gamma(\beta + \alpha   k)}\) denotes a Mittag-Leffler function, \(\mathcal{E}^\prime\) its derivative, and the \(I_k\) a modified Bessel function. The dimensionless gamma constants are \(\gamma_s = \beta_s/(\beta_l+\beta_r)\).

\subsection{Integrals and identities} \label{Appendix:Integrals}

\textbf{Exponential integrals:} The following integral identity is extremely useful both in calculating the partial moments of energy, and with calculating the leading order power and number rates for the smart demons,
\begin{align*}
\int_{E_0}^\infty \mathbf{d}E \, E^p e^{-\beta E} = \frac{\Gamma(p+1, \beta E_0)}{\beta^{p+1}} = \frac{p!}{\beta^{p+1}}e^{-\beta E_0} \sum_{j=0}^p \frac{(\beta E_0)^j}{j!}.
\end{align*}
For full moments (\(E_0 = 0\)), this reduces to
\begin{align*}
\int_0^\infty E^p e^{-\beta  E} = \frac{p!}{\beta^{p+1}}  = \frac{\Gamma(p+1)}{\beta^{p+1}} \text{ : \(p\ge 0\)}
\end{align*}

\textbf{Some properties of the incomplete gamma function:} The incomplete gamma function is defined to be 
\[
\Gamma (s, z) = \int_z^\infty t^{s-1} e^{-t} \mathbf{d}t.
\]
As a consequence, we have the recursive formula \(\Gamma(s+1, z) = s \Gamma(s, z) + z^s e^{-z} \), and the special cases \(\Gamma(s,0) = \Gamma(s)\) and \(\Gamma(1, z) = e^{-z} \). For integer values of \(s\), this recursion can be expanded to give us
\begin{align*}
\Gamma(s+1, z) = e^{-z} \sum_{k=0}^s \frac{s!}{(s-k)!} z^{s-k}.
\end{align*}

One integrals that occurs that involves the incomplete gamma function is 
\begin{align*}
    \int_0^\infty \mathbf{d}E \, e^{-\beta_1 E} \, E^k \, \Gamma(n, \beta_2 E) = \frac{\Gamma(n+k+1)}{(k+1) \beta_2^{k+1}} F_{2,1}\left(k+1, n+k+1; k+2; -\frac{\beta_1}{\beta_2} \right)
\end{align*}
which is useful in deriving (\ref{PowerDemonCurrents}).

\subsection{Demon energy and number currents for \(d>1\)} \label{Appendix:HigherDimensionDemons}

When  \(\kappa_l\) and \(\kappa_r\) are very small, both the energy and number demons act like the direction demon, since they always let particles pass from left to right if no particles are passing from right to left, and the probability that particles pass from right to left \emph{and} from left to right during a time window becomes very small.

Because of this, we can expand the energy and number currents of the energy and number demons as the direction demon current, plus terms of increasing order in \(\kappa_l\) or \(\kappa_r\) that correspond to events where the demon opens the door even though some particles are passing from right to left. We will treat \(\kappa_l\) and \(\kappa_r\) as having the same order.

For the energy demon, the next order term for \(P_\tau^{(e)}\) after the direction demon term should be have a factor of \(\kappa^2\). The only relevant event is where one particle approaches from each side, and \(E_l>E_r\). The event where two particles approach from the left, but none from the right is already included in the direction demon term, and the event where two particles approach from the left, but none approach from the right always has \(E_r > E_l = 0\), so the door will not open and allow this event.

We integrate the probabilities from (\ref{ProbabilityNE}) over allowed energies, with \(n_l = 2\), \(n_r = 1\), to find the leading correction for \(P_\tau^{(e)}\)
\begin{align*}
    \frac{1}{\tau} \int_0^{\infty} \mathbf{d}E_r \, \int_{E_r}^\infty \mathbf{d}E_l 
    \frac{\kappa_l (\beta_l E_l)^{D} }{\Gamma(D)} \frac{e^{-\beta_l E_l-\kappa_l}}{E_l}
    \frac{\kappa_r (\beta_r E_r)^{D} }{\Gamma(D)} \frac{e^{-\beta_r E_r-\kappa}}{E_r} \times (E_l - E_r)
\end{align*}
with the result being the correction term of \(P^{(e)}_\tau\) from (\ref{PowerDemonCurrents}).

To evaluate the number current for the energy demon, we first notice that there are no relevant terms of order \(\kappa^2\). The event that was the most significant for the power does not count towards the number current since \(n_l-n_r=0\). The next relevant term is from the events \(n_l=2\), \(n_r=1\), and \(n_l=1\), \(n_r=2\). We must still integrate this over the space of energies with \(E_l > E_r\).
\begin{align*}
\frac{1}{\tau} \int_0^{\infty} \mathbf{d}E_r \, \int_{E_r}^\infty \mathbf{d}E_l 
    \left( \frac{\kappa_l^2 (\beta_l E_l)^{2D} }{\Gamma(2D)2!} \frac{e^{-\beta_l E_l-\kappa_l}}{E_l} \frac{\kappa_r (\beta_r E_r)^{D} }{\Gamma(D)} \frac{e^{-\beta_r E_r-\kappa_r}}{E_r}  - \frac{\kappa_r^2 (\beta_r E_r)^{2D} }{\Gamma(2D)2!} \frac{e^{-\beta_r E_r-\kappa_r}}{E_r} \frac{\kappa_l (\beta_l E_l)^{D} }{\Gamma(D)} \frac{e^{-\beta_l E_l-\kappa_l}}{E_r}\right)
\end{align*}
The first term is from \(n_l=2\), \(n_r=1\), the second is from \(n_l=1\), \(n_r=2\). The result is the correction term of \(N^{(e)}_\tau\) from (\ref{PowerDemonCurrents}).

For the number demon, the leading order event is \(n_l=2\), \(n_r=1\) since the events \(n_l=0\), \(n_r=2\), and \(n_l=1\), \(n_r=1\), and  \(n_l=0\), \(n_r=2\) do not satisfy \(n_l > n_r\). Using (\ref{ProbabilityOfNumber}), it is easy to see that
\begin{align*}
    \frac{1}{\tau} \cdot \frac{\kappa_l^2}{2!} \kappa_r\,e^{-\kappa_l-\kappa_r},
\end{align*}
is just the correction term for \(I^{(n)}_\tau\) in (\ref{NumberDemonCurrents}). To find the correction term for the number demon's power, we integrate (\ref{ProbabilityNE}) over all energies,
\begin{align*}
     \frac{1}{\tau} \int_0^\infty \mathbf{d}E_r \, \int_0^\infty \mathbf{d}E_l 
    \frac{\kappa_l^2 (\beta_l E_l)^{2D} }{\Gamma(2D)2!} \frac{e^{-\beta_l E_l-\kappa_l}}{E_l} \frac{\kappa_r (\beta_r E_r)^{D} }{\Gamma(D)} \frac{e^{-\beta_r E_r-\kappa}}{E_r} \times (E_l - E_r),
\end{align*}
obtaining the correction term of \(P^{(n)}_\tau\) in (\ref{NumberDemonCurrents}).

\subsection{Demon entropy production} \label{Appendix:DemonEntropy}

\textbf{Demon entropy production.} While the action of the demon reduces the entropy of the system, the operation of the demon must itself result in an increase in entropy, \(S_\mathrm{dem}\). By gathering information about the system, the full phase space is reduced into a subset - the phase space \emph{given} the measurement outcome. Since the demon operates cyclically, it must erase all all the information it has gained via measurement, which must be at least \(H[\hat{M}]\). Therefore, it produces entropy at a rate,
\begin{align}
\dot{S}_\mathrm{dem} \cdot \tau / k_B \ge H[\hat{M}] &= H[\hat{X}] - H[\hat{X} \vert \hat{M}].
\label{DemonEntropy}
\end{align}
where the random variables \(\hat{X}\) denotes the system state and \(\hat{M}\) a measurement of sub-state necessary to make a decision. 

The fact that \(H[\hat{M}] = H[\hat{X}] - H[\hat{X} \vert \hat{M}]\) is a consequence of Bayes law for conditional entropy, \(H[\hat{M} \vert \hat{X}] = H [ \hat{X} \vert \hat{M}] - H [\hat{X}] + H[\hat{M}]\), and that fact that \(H[\hat{M} \vert \hat{X}] = 0\) since the state of the system will completely determine the measurement \(\hat{M}\) for all our demons.

\textbf{Calculation of demon entropy production.} Here we detail how to use (\ref{DemonEntropy}) to calculate the entropy production of the demons. \added{Recall that} \(\hat{X}\) is the state of the system (a random variable), and \(\hat{Y}\) is the demon measurement (also a random variable) that will vary from demon to demon. We will use the energy demon as our example, the other demons are analogous.

The total entropy of the system is just the entropy of the particle and number distribution,
\begin{align*}
H[\hat{X}] = S[p^{(l)} p^{(r)}] = -\sum_{n_l,n_r \ge 0} \int_0^\infty \mathbf{d}E_l\,\mathbf{d}E_r\, p^{(l)}_{n_l}(E_l) p^{(r)}_{n_r}(E_r) \log \left[ p^{(l)}_{n_l}(E_l) p^{(r)}_{n_r}(E_r)  \right].
\end{align*}
The measurement, \(\hat{Y}\), for the energy demon is the function \(\hat{Y}(\omega) = 0\) if \(E_l \le E_r\) for outcome \(\omega\), and \(\hat{Y}(\omega) = 1\) if \(E_l > E_r\). Consequently, let 
\begin{align*}
    & P_0(n_l, n_r, E_l, E_r) = p_{n_l}(E_l) p_{n_r}(E_r) \Theta(E_r - E_l) / p_0
    \qquad 
    P_1(n_l, n_r, E_l, E_r) = p_{n_l}(E_l) p_{n_r}(E_r) \Theta(E_l - E_r) / p_1
    \\
    & p_0 = \sum_{n_l, n_r\ge 0}  \int_0^\infty \mathbf{d}E_l\,\mathbf{d}E_r p_{n_l}(E_l) p_{n_r}(E_r) \Theta(E_r - E_l)
    \qquad 
    p_1 = 1-p_0.
\end{align*}
That is, \(P_0\) and \(P_1\) are just the probability distribution \emph{given} that \(E_l < E_r\) or vice-versa, and \(p_0\), \(p_1\) are the probabilities that \(E_l < E_r\) or vice-versa.

The conditional entropy is just
\begin{align*}
    H [ \hat{X} \vert \hat{Y} ] = p_0\cdot S(P_0) + p_1 \cdot S(P_1).
\end{align*}

For other demons, \(H[\hat{X}]\) has the same value, the only different part is calculating the conditional probabilities \(P_0\), \(P_1\).

\begin{figure}[t]
\centering
\includegraphics[width=0.6\textwidth]{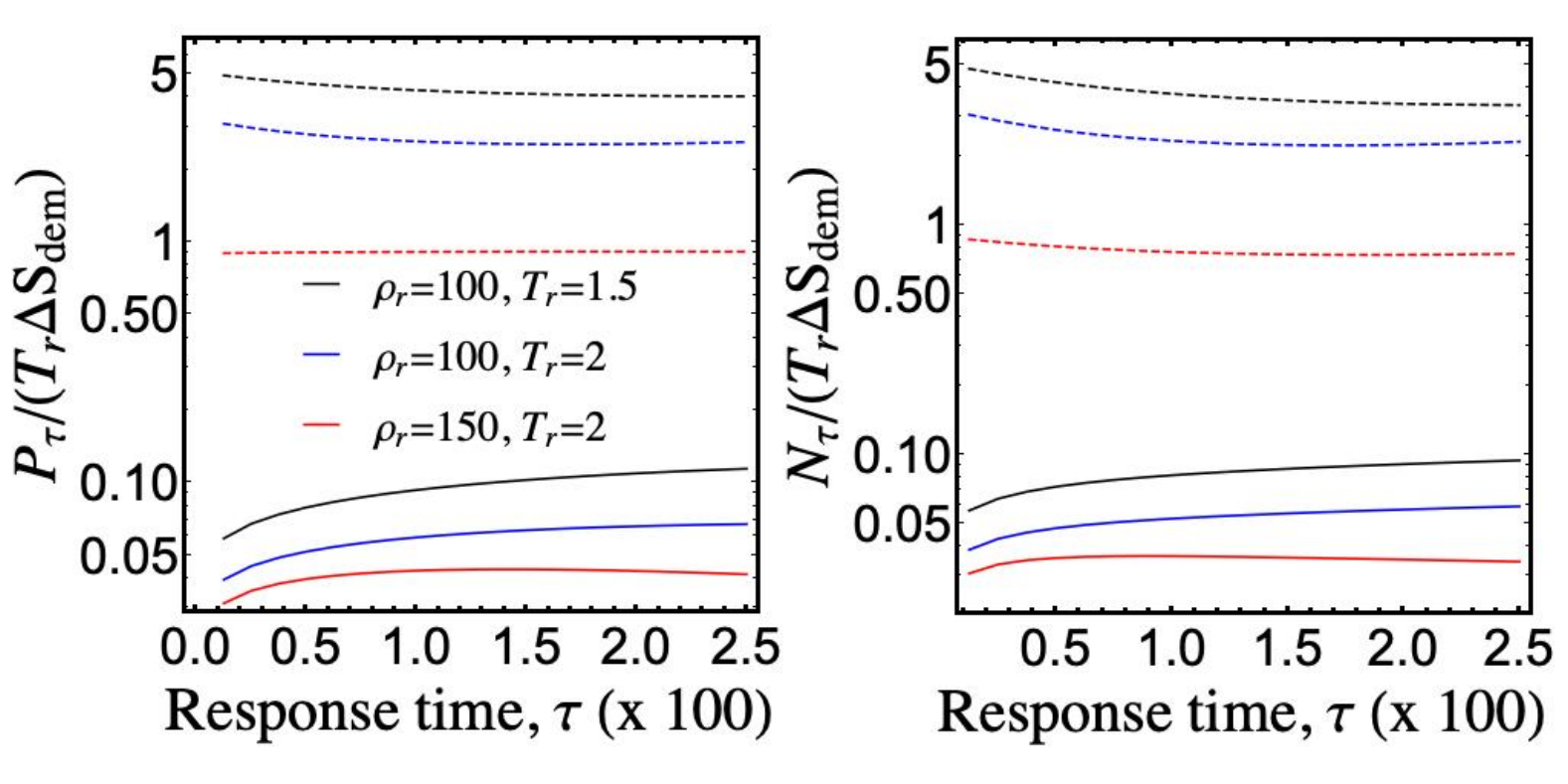}
\caption{
    \textbf{The effect of spatiotemporal restrictions on coefficients of performance.} We plot the ratio of heat transferred to heat generated (left) and mass transferred to heat generated (right) for several sets of system parameters, for our energy demon (solid) and an ideal heat / mass pumps (dashed). The demons generate entropy, whereas the ideal pumps do not. Length and time restrictions lead to two orders of magnitude of  reduction in the coefficient of performance for the indicated density and temperatures. The left subsystem has \(\rho_l = 100\), \(\beta_l = 1\).
}
\label{Fig:efficiency}
\end{figure}

To measure the efficiency of the power demon, we look at the ratio of power or number current to heat generated by the demon, and compare these values to their theoretical minimum values (when \(\dot{S}_{\text{dem}} = \dot{S}_r + \dot{S}_l\)) (Fig. \ref{Fig:efficiency}). As expected, the entropy generated by the demon surpasses the reduction in system entropy due to the demon's activities, resulting in suboptimal efficiencies at all values of \(\tau\).

\subsection{Simulation details} \label{Appendix:SimulationDetails}

\textbf{One dimensional demons.} One dimensional simulations are run in Mathematica. A system with 1,000,000 particles on each side is created. The size of the sides is scaled to match the requested number densities, \(\rho_l\), \(\rho_r\), and the particles are give \(x\) coordinates on their sides uniformly at random. The demon gate is thought of as being at \(x=0\). Each particle has its velocity initialized according to the Gibbs distribution, \(v \sim \mathcal{N}(0, 1/\sqrt{\beta\,m})\). The times at which particles hit the demon door, which only depends on their velocity and initial position (since the particles are non-interacting) are recorded.

As we do not enforce particles to stay within bounds via walls, we only sample from events within a short enough time that only particles that start near the gate can possibly pass through the gate, and particles near the gate will not be able to reach the far wall. The hallmark of this behavior is that the rate of events is linear in time. Empirically, we have found that keeping the events in only the first \(T = \frac{N}{40 \max (\rho_l, \rho_r)}\) amount of time (for the temperatures we probe) is more than sufficient to be safe. 

Once we have our list of gate passing events, we can bin time into bins of various \(\tau\) and check what the resulting currents would be if the demon operated with this \added{reaction time}. An energy demon, for example, would only count particle and energy contributions from time windows where the net energy of particles coming from the left is greater than the net energy of particles coming from the right.

\textbf{Two dimensional demons.} The two dimensional demon is simulated using the GFlow molecular dynamics package \footnote{Code available at \url{https://github.com/nrupprecht/GFlow}}. The simulation consists of a box of dimensions \(2 W \times W\), with a vertical wall consisting of hard sphere particles at \(x=0\), separating the system into two sides, and a gap in the wall with a special gate of length \(L\) that comprises the demon door.

Out of computational necessity, the gas particles have finite size, which allows them to interact with the wall and demon gate. Strictly speaking, this makes the system non-ideal. When the door is ``open,'' the particles that it consists of are simply moved out of the simulation space, allowing for particles to pass through the opening uninhibited. While the door is closed, the particles form a wall, blocking the hole, and act as hard spheres. Act of door closing must be treated more carefully. In the ideal gas case, the particles would be point particles, so the demon door would never hit the gas particles when it is closing. However, we need a strategy to deal with what happens if the door needs to close while a particle is in the way. Simply placing the particles back in their positions can result in gas particles overlapping with the door, resulting in a huge addition of energy to the system. To solve this, when we close the door, we place all the door particles back into position, and teleport any particles that overlap with the door to a random position on their side where they do not overlap with any particles (and keeping their velocity constant).

To simulate the demon, we keep track of what side each particle is on at the start of each time window, and then let the simulation run for \(\tau\) with the demon's door open. At the end of \(\tau\), we count the amount of energy and particles that have passes from left to right, and from right to left. If these quantities are acceptable to the demon we are simulating, we repeat the process for the next \(\tau\). If not - for example, if net energy flux was to the left, and the demon is an energy demon - we revert the simulation back the the start of \(\tau\) (we have saved the particles' positions and velocities), close the demon's door, and run for \(\tau\). We repeat this process for the requested amount of time, keeping track of the number and energy currents during each time window.

Though, as seen in Fig \ref{Fig:Demon2D}, the agreement between simulation and theory is quite good (especially for the demons working with ideal gas, which is what our derivation is for), there are a number of finite size effects that can skew our results. One of the most noticeable effects is the skewing of results for very small \(\tau\). Because of the teleportation that can occur when the door closes with a particle in the way (which in turn can occur because of the finite size of the particles), the door rapidly opening and closing can have the effect of removing particles very near (touching) the door. Depending on the number densities and temperatures of the subsystems, we have observed this effect skewing the rates to be either larger or smaller than the prediction. Even so, the predicted values are still fairly close to the actual values observed in simulations.

Another finite size effect is the finite size of the system. As particles pass from one side to another, the number density and temperatures of the sides change. We solve this by generating large enough systems that the change in subsystem parameters is only a few percent over the entire course of the simulation. However, it is still possible for these effects to skew the data slightly.

We close this section by commenting on how the demon changes when interactions are added. The degree to which (short range) interactions have an effect on the system varies based on the packing fraction of the system, \(\phi_s = N_s v / V_s = \rho_s v\) where \(v\) is the volume of a single particle. We can think of how the energy and number currents of a system with fixed \(\rho_l,\,\rho_r,\,\beta_l,\,\beta_r\) changes as \(v\) is increased. In the case of hard spheres, an event that becomes common as \(v\) increases is that a particle passes through the gate, hits a particle on the other side, and bounces back to the original side. From the point of view of the demon, which only counts the net changes that occur during each \(\tau\), energy is transfered from one side to another by this event without particle transfer.

Other particle types, such as Lennard Jones particles, are possible to simulate, and could have other interesting behaviors, like the tendancy of LJ particles to clump together. For potentials like LJ particles, where potential energy is a significant part of the total energy, the demon may want to take a different strategy, like counting the potential energy of clusters of particles that pass through the gate to make its decision, as opposed to just the kinetic energy.

\bibliographystyle{apsrev4-1}
\bibliography{bibliography.bib}

\end{document}